\documentclass{mn2e}

\newcommand{\msunyr}{\mbox{\,{\rm M}$_\odot$\ {\rm yr}$^{-1}$\,}}
\newcommand{\msun}{\mbox{\,{\rm M}$_\odot$\,}}
\newcommand{\kms}{\mbox{${\,\rm km~s}^{-1}$}\,}
\newcommand{\cmsq}{\mbox{${\,\rm cm}^{-2}$}\,}
\newcommand{\ergs}{\mbox{${\,\rm erg~s}^{-1}$}\,}
\newcommand{\etal}{{et al.}}

\begin{document}

\title[The Cluster Wind from Local Massive Star Clusters] 
{The Cluster Wind from Local Massive Star Clusters} 
 
\author[I.R. Stevens, J.M. Hartwell]
{Ian R. Stevens, Joanna M. Hartwell\\
School of Physics and Astronomy, 
University of Birmingham, Edgbaston, Birmingham B15 2TT \\
(E-mail: irs@star.sr.bham.ac.uk, jmh@star.sr.bham.ac.uk)}

\maketitle

\begin{abstract}

Results of a study of the theoretically predicted and observed X-ray
properties of local massive star clusters are presented, with a focus on
understanding the mass and energy flow from these clusters into the ISM
via a cluster wind. A simple theoretical model, based on the work of
Chevalier \& Clegg (1985), is used to predict the theoretical cluster
properties, and these are compared to those obtained from recent {\sl
Chandra} observations. The model includes the effect of lower energy
transfer efficiency and mass-loading.  In spite of limited statistics,
some general trends are indicated; the observed temperature of the
diffuse X-ray emission is lower than that predicted from the stellar
mass and energy input rates, but the predicted scaling of X-ray
luminosity with cluster parameters is seen. The implications of these
results are discussed.

\end{abstract}


\section{Introduction} \label{sect:intro}

Super Star Clusters are dense clusters of young massive stars, first
identified in NGC\,1275 by Holtzman \etal\ (1992) using the {\sl Hubble
Space Telescope} (HST), and subsequently in a wide range of star-forming
galaxies, such as merging systems (NGC\,4038/4039; Whitmore \& Schweizer
1995), dwarf galaxies (Henize 2-10; Johnson \etal\ 2000), classical
starbursts (M82; Gallagher \& Smith 1999) amongst many other systems
(see Whitmore 2000 for a review).  These extragalactic star clusters can
contain many thousands of very young, energetic stars, and have stellar
densities far greater than those seen in normal OB associations. The
values quoted by Whitmore (2000) indicate ages for these star clusters
of typically $1-10$Myr, radii typically in the range of $\sim 1-6$pc,
total masses of stars in the cluster in the range $10^3- 10^6\msun$),
with the central stellar densities reaching up to $\sim
10^5\msun$~pc$^3$. It is clear that in many galaxies a substantial
fraction of the ongoing star-formation (and hence mass/energy injection
into the ISM) is occurring in SSCs (Origlia \etal\ 2001).  The Galactic
(or local analogues) of these extragalactic SSCs are objects such as
NGC\,3603, R136 in 30 Doradus and the Arches cluster near the Galactic
Centre (see Figer, McLean \& Morris 1999b).

The component stars of an SSC are believed to be roughly coeval, and
when the cluster is very young the massive stars in the cluster will
have strong stellar winds which will combine to produce an outflow from
the cluster, which is referred to as the cluster wind. As the cluster
ages mass and energy input from supernova explosions will begin to
dominate (cf. Leitherer \& Heckman 1995).  In many starburst galaxies,
where a large number of SSCs are seen, this cluster wind will be an
important contributor to energising the interstellar medium and possibly
driving a hot superwind or outflow, such as those seen in M82 or
NGC\,253 (Lehnert, Heckman \& Weaver 1999; Strickland \etal\ 2002). The
efficiency with which stellar kinetic energy (from both stellar winds
and supernovae) is converted into thermal energy within the cluster,
which in turn can drive an outflow, is an important and very difficult
parameter to determine. In the literature, several different values or
prescriptions for the thermalization efficiency have been used, often
encapsulated in a parameter $\eta$, which represents the fraction of the
kinetic energy of stars and supernova in the cluster that is
thermalized. The fraction of the kinetic energy that is not thermalized
is assumed to be lost to the system, primarily via  radiative losses.
As we shall see radiative losses at X-ray energies are
likely to very small, however, at UV and IR energies radiative
losses are likely to be much larger. In fact, inhomogeneities in the
complex flow within the cluster could easily lead to the formation of
denser regions that could contribute to radiating energy away from
the system. 

In the literature, Strickland \& Stevens (2000) argued for very
efficient thermalization ($\eta\sim 1$; see also Chevalier \& Clegg
1985), whereas Bradamante, Matteucci \& D'Ercole (1998) argued for a
much lower thermalization efficiency of a few per cent. Other, more
complex, prescriptions of $\eta$ have also been proposed. For instance,
Recchi, Matteucci \& D'Ercole (2001) use a model with $\eta$ being very
low in the early evolution of a cluster, but as material is evacuated by
the cluster wind to create a low density region the thermalization
efficiency rises to close to unity. In relation to LMC superbubbles, Oey
(1996) noted that the bubble dynamics implied a relatively low
thermalization efficiency (or put another way that there was less power
driving the bubbles than implied by the stellar populations within the
bubbles, see also Chu \& Mac Low for a discussion of the X-ray
properties of superbubbles). It would clearly be very useful to have
more direct observational constraints on this process.

Understanding the mass and energy loss processes from distant SSCs via
the hot cluster wind can best be done at X-ray energies. However,
because of their distance and the high stellar density, even {\sl
Chandra} has insufficient resolution to see what is going on in and
around these systems, and resolve the diffuse emission from point source
emission from stars in the cluster. However, local lower-mass analogues
of these SSCs in our Galaxy or Local Group can be used to study the mass
and energy loss processes from stellar clusters.  By studying the
diffuse X-ray properties of star clusters and the stellar energy
injection rate from stellar populations in the cluster it should be
possible to constrain the efficiency with which the clusters can convert
kinetic energy to thermal energy. Although not perfect analogues, the
values determined could have far-reaching consequences, not just for the
understanding of the processes within the star clusters, but also for
the energy emission from larger structures such as starburst galaxies,
where many SSCs are present.

It is also worth making the point that while the mass and energy
injection rate from massive stars via their stellar winds can be
estimated to a reasonable degree of accuracy, the same is less true for
supernova. For instance, it is worth noting that the type IIn supernova
SN1988Z is believed to have radiated $\sim 10^{52}$~erg in its early
evolution (Aretxaga \etal\ 1999), see also Chevalier \& Fransson 2001),
which is rather more than the usually assumed values for the total
energy injection from a SN. Consequently, looking at very young
clusters, where SN injection does not dominate, has considerably more
promise.

In this paper the theory of cluster winds is developed, particularly as
it relates to X-ray emission from clusters (\S~2), and it is then
applied to results from {\sl Chandra} observations of nearby SSCs or
smaller stellar clusters (\S~3). The results are discussed in \S~4 and
summarised in \S~5.\\

\begin{figure*} 
\vspace{13cm} 
\includegraphics{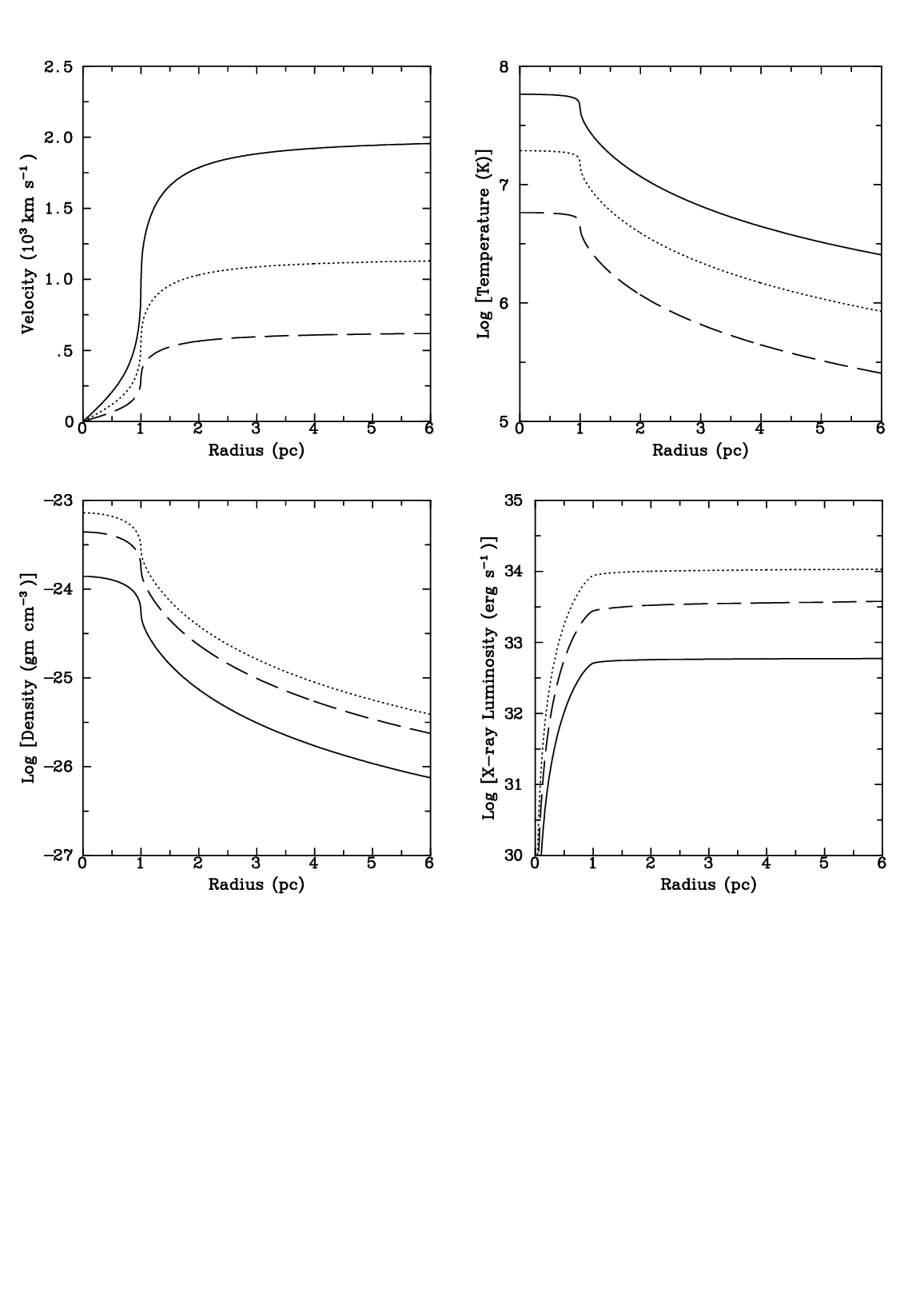} 
\caption{The radial structure for the cluster wind, as calculated from
the Chevalier \& Clegg (1985) model (see text for details). Shown are
the radial profiles for the wind velocity (top left panel), gas
temperature (top right), density (bottom left) and the cumulative X-ray
luminosity (bottom right). The standard model (solid line) is for a star
cluster with a stellar mass injection rate of $\dot
M_\ast=10^{-4}\msunyr$, ${\bar V}_\ast=2000\kms$, $\eta=1$ and no
additional mass-loss ($\dot M_{cold}=0$). The second model is identical,
except with $\eta=0.1$ (dashed line), and the third model (dotted line)
has $\eta=1$ and $\dot M_{cold}=2\times 10^{-4}\msunyr$.}
\label{fig1}
\end{figure*}

\section{The X-ray Emission from a Cluster Wind}

When massive stars with strong winds are in close proximity, such as in
a stellar cluster, the winds collide both with each other and with the
surrounding ISM, filling the surrounding volume with hot shocked
gas. Eventually all of the available volume is filled and the hot gas
escapes beyond the boundaries of the cluster. The development of such a
cluster wind has been numerically modelled by Cant\'o, Raga \&
Rodr\'iguez (2000) in a simulation carried out assuming a spherical
cluster consisting of $\sim 30$ stars (see also Ozernoy, Genzel \& Usov
1997; Raga \etal\ 2001).

The wind from a stellar cluster can also be described using the
Chevalier \& Clegg (1985) model originally used to describe outflows
from starburst galaxies. The contrast between these models is that
Cant\'o \etal\ (2000) also used hydrodynamic calculations with mass and
energy input from discrete stars within the cluster, whereas the
Chevalier \& Clegg (1985) model assumed mass and energy injection
uniformly throughout the starburst region. The dense star clusters
considered here do not have a uniform distribution of stars and tend to
be centrally concentrated, and mass and energy injection will tend to
follow the stellar distribution. However, Cant\'o \etal\ (2000), who did
include a non-uniform stellar distribution showed that their numerical
simulations generally reproduce the results of the Chevalier \& Clegg
type model, and the formulation of Chevalier \& Clegg (1985) is adopted
here. One important difference between the Cant\'o \etal\ (2000) and
Chevalier \& Clegg (1985) models is that the Cant\'o \etal\ (2000) have
a much greater range of temperatures (and densities inside the cluster
core).

If mass and energy from stars is being injected via stellar winds into
the volume of a stellar cluster of core radius $R_c$ then Chevalier \&
Clegg (1985) showed that the solution for the outflow from this region
can be written as

\begin{equation}
\left(\frac{3\gamma+1/{\cal M}^2}{1+3\gamma}\right)
^{\textstyle-\frac{(3\gamma+1)}{(5\gamma+1)}}
\left(\frac{\gamma-1+2/{\cal M}^2}{1+\gamma}\right)
^{\textstyle\frac{(\gamma+1)}{2(5\gamma+1)}}=\frac{r}{R_c}
\label{eqn1}
\end{equation}

\noindent for $r<R_c$ and

\begin{equation}
{{\cal M}}^{\left(2/(\gamma-1)\right)}
\left(\frac{\gamma-1+2/{\cal M}^2}{1+\gamma}
\right)^{\left((\gamma+1)/2(\gamma-1)\right)}= 
\left(\frac{r}{R_c}\right)^2
\label{eqn2}
\end{equation}

\noindent for $r>R_c$, where ${\cal M}$ is the Mach number of the flow,
$\gamma$ is the adiabatic index (with $\gamma=5/3$ assumed hereafter),
$R_c$ the radius of the star cluster and $r$ is the radius from the
center of star cluster. From these equations and the mass and energy
injection rates, the velocity $v(r)$, temperature $T(r)$ and mass
density $\rho(r)$ can be determined. The technique to do so involves
solving eqn.~\ref{eqn1} or eqn.~\ref{eqn2} as appropriate, for the Mach
number at each radius, using the Newton-Raphson method. From the Mach
number, and using the integrated forms of the mass and energy energy
equations, the other variables can be determined (see Chevalier \& Clegg
1985 for more details). Solar abundances and fully ionized material with
a mean mass per particle of ${\bar\mu}=10^{-24}$~gm are also assumed.

It is worth noting that the conditions in the central regions of the
cluster are a good indicator of the recent history of the cluster. The
flow time within the cluster is short, typically a few $10^3$~yr,
whereas the characteristic timescale of the larger scale bubble being
blown by the cluster is much longer (i.e. of order the age of the
cluster). This means that the current X-ray properties should give a
better handle on the transfer efficiency of stellar wind energy to the
cluster wind.

\subsection{The Stellar Mass and Energy Injection Rate}

For the young star clusters under consideration here the energy
injection is dominated by stellar winds, especially those of the massive
OB and Wolf-Rayet (WR) stars, and we assume that no supernovae have as
yet exploded and contributed to the cluster wind. At later epochs
supernovae begin to dominate the energy injection (Leitherer \& Heckman
1995).

In principal, from the current observed massive star population in the
clusters, the mass and energy injection rate from the stellar winds of
these stars can be estimated, and this is the procedure we
adopt. However, as discussed elsewhere, there are some significant
uncertainties in the determination of these values, related to
uncertainties in understanding the structure and evolution of massive
stars. 

The bulk of the mass and energy injection is assumed to happen within a
cluster core radius $R_c$.  If the cluster contains $N_\ast$ stars, each
star with a mass-loss rate of $\dot M_i$ and wind terminal velocity
$V_i$ with $i=1\dots N_\ast$ then the total stellar wind mass injection
rate, $\dot M_\ast$, is

\begin{equation}
\dot M_\ast=\sum_i^{N_\ast} \dot M_i
\label{eqn3}
\end{equation}

\noindent and the total kinetic energy injection rate is

\begin{equation}
\dot E_\ast=\sum_i^{N_\ast} \frac{1}{2}\dot M_i V_i^2 \ .
\end{equation}

\noindent A mean weighted terminal velocity for the stars in the
clusters is defined such that

\begin{equation}
{{\bar V}_\ast}= \left(\frac{\sum_i^{N_\ast} \dot M_i V_i^2}{\dot M_\ast}
\right)^{1/2}
\label{eqn5}
\end{equation}

For some of the stellar clusters discussed in the next section the
individual mass-loss rates and terminal velocities can be determined,
and for those systems where there is insufficient observational evidence
stellar evolution models, such as the {\sl Starburst99} code (Leitherer
\etal\ 1999) can be used. 

Some qualifications should be made here. The method used for estimating
the mass-loss rates and terminal velocities of the cluster stars rests on
our current best understanding of the mass-loss parameters of massive
stars. There remain significant deficiencies in our knowledge of the
structure and evolution of massive stars and their winds.  For instance,
the {\sl Starburst99} code only assumes a population of single stars and
binary evolution could make a difference to the mass-loss rates and
terminal velocity. Indeed, it is known that the characteristics of the
winds of massive stars evolve with time, and this snapshot could be
missing important early phases of evolution when the winds were much
stronger. In spite of these qualifications we shall use current best
estimates and proceed accordingly.

In a cluster although the optically visible stars (and their winds) will
likely dominate the energy injection, these stars will not necessarily
be the only sources of mass. Protostars (or other optically obscured
objects) will also contribute mass as material is ablated off them by
the cluster wind or through radiation. Examples of such objects may be
the ProPlyDs which have been seen in NGC\,3603 and elsewhere (for
instance, Brandner \etal\ 2000; M\"ucke \etal\ 2002).  This cold mass
being injected into the cluster wind can be accounted for by a term
$\dot M_{cold}$, so that the total mass-loss rate into the cluster is

\begin{equation}
\dot M_{tot}=\dot M_\ast + \dot M_{cold}
\end{equation}

\noindent This process can be considered as a mass-loading process (see
Hartquist \etal\ 1997 and references therein).

As discussed in the introduction, to account for the effects of
radiative losses in the conversion of stellar wind energy into the
cluster wind, we can introduce a parameter that we term the \lq\lq
thermalization efficiency\rq\rq\  parameter $\eta$, such that the energy
available to drive the cluster wind is $E_{th}=\eta E_\ast$, so that

\begin{equation}
\dot E_{th}=\eta \sum_i^{N_\ast} \frac{1}{2}\dot M_i V_i^2 =
\frac{\eta}{2}{\dot M_\ast} {{\bar V}_\ast}^2
\end{equation}

\noindent and allow $\eta$ to vary between 0 and 1. This parameter
allows for radiative losses. The energy radiated at X-ray energies is
unlikely to be a major loss mechanism, but UV radiation, particularly
from interfaces between hot and cold gas, may be a bigger contributor,
as could IR radiation.  As discussed by Recchi \etal\ (2001) we might
expect this parameter to be relatively small in the early stages of
cluster evolution, but to increase at later times. For simplicity, in
these simulations $\eta$ is assumed constant. The result of 
allowing $\eta$ to be less than unity and additional \lq\lq cold\rq\rq\
mass-injection will be to reduce the terminal cluster wind velocity
below ${\bar V}_\ast$. This is turn increases the density and often
leads to an enhanced X-ray luminosity.

The assumption of a constant thermalization efficiency is also highly
simplistic. Not only is the value likely to vary as a function of time,
as the cluster wind evolves, it is also likely to vary as a function of
location within the cluster, with a different thermalization efficiency
in the denser inner regions to that in the more diffuse outer
regions. Given the lack of detailed information as to how $\eta$ might
vary, we shall treat it as a constant. So, subject to this recognition
that this is a highly simplified treatment of a very complex problem we
shall proceed.

\subsection{The Cluster X-ray Luminosity}

Observationally, at X-ray wavelengths, a young stellar cluster will have
both point source emission associated with individual stars, such as
from colliding winds in massive binaries, intrinsic emission from single
early-type stars, emission associated with SNRs, X-ray binaries or
pre-main sequence stars, and diffuse emission (due to the cluster
wind). Here we concentrate only on the diffuse X-ray emission, and only
from younger clusters where we would not expect to see emission from
SNRs and X-ray binaries. In the data presented later, the point sources
have been removed from the X-ray data to leave only the diffuse
emission. It is however possible that unresolved point sources,
particularly from lower luminosity pre-main sequence (PMS) stars, will
contaminate the diffuse emission. We will discuss this point more later.

For a given model, the broad-band X-ray luminosity of the cluster will
be given by

\begin{equation}
L_X=\int_0^{R_c}  4\pi r^2 n_e n_i \Lambda(T) dr
\label{eqn8}
\end{equation}

\noindent where $\Lambda(T)$ is the emissivity of gas at a temperature
$T$, and $n_e$ and $n_i$ are the electron and ion number density
respectively.  Given the solution for $\rho(r)$, $v(r)$ etc,
eqn.~\ref{eqn8} can be evaluated to derive the total X-ray luminosity
(over a wide waveband) of the cluster. A more detailed analysis could
calculate the luminosity in specific wavebands. However, the majority of
the emission will fall in the {\sl Chandra} waveband and we will not
complicate matters further.

In Fig.~\ref{fig1} the radial variation of the velocity and temperature
for a stellar cluster are shown, where the total stellar mass-loss rate
is $10^{-4}\msunyr$ and ${\bar V}_\ast=2000\kms$ and the cluster core
radius is 1pc. These values are similar to those of the clusters
discussed in this paper. In this model the central temperature is
$5.8\times 10^7$K, the central ion number density is $0.67$cm$^{-3}$,
and the X-ray luminosity of the region within a radius $R_c$ is
$L_X=5.1\times 10^{32}\ergs$.  Importantly for our purposes, the cluster
itself dominates the X-ray emission and the volume outside $R_c$
contributes only 15 per cent of the total X-ray luminosity. The reason
for this is the sharp fall off in density (and temperature) outside the
cluster, which more than compensates for the increasing volume of the
emission region.  For a simple analysis such as presented here this
level of accuracy is sufficient.

The central temperature of the cluster wind ($T_0$) in this model is
given by

\begin{equation}
\left(\frac{T_0}{\rm{K}}\right) =
1.45\times10^7\left(\frac{{\bar V}_\ast}
{1000\kms}\right)^2\ .
\label{eqn9} 
\end{equation}

\noindent Note that the slightly different numerical constant as
compared to Cant\'o \etal\ (2000) is due to slightly different
assumptions as to the mean mass per particle.

From the radial solution the emission weighted temperature for the
cluster region ($T_{cl}$) can also be calculated, which will be somewhat
lower than the central temperature. For instance, for the standard model
with ${\bar V}_\ast=2000\kms$, the emission weighted temperature is
$T_{cl}=5.1\times 10^{7}$~K.

\begin{figure}
\vspace{7cm}
\includegraphics{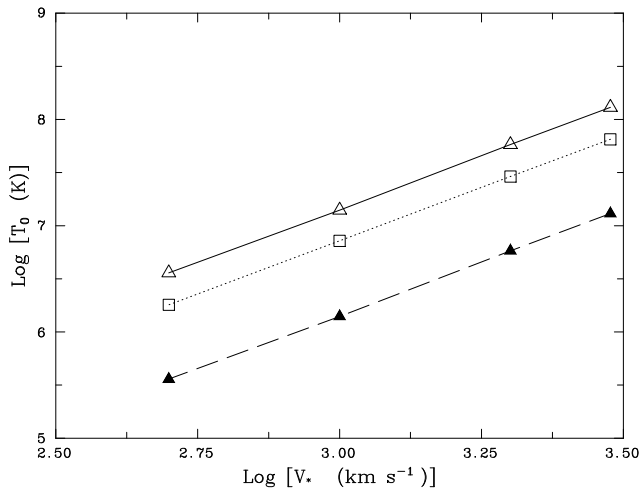}
\caption{The variation of the expected X-ray temperature of the cluster
core versus the wind velocity of the stellar component.  Shown are model
results for the cluster central temperature and cluster luminosity for
varying $\eta$ and the contribution of cold gas. The standard cluster
model has $\dot M_\ast=10^{-4}\msunyr$ and $R_c=1$pc. The models
represented by open triangles and marked with a solid line are for
$\eta=1$ and ${\bar V}_\ast=500, 1000, 2000$ and $3000\kms$ The models
represented by filled triangles and marked with a dashed line are the
same models except with $\eta=0.1$.  The models represented by open
squares and marked with a dotted line are the same models except with
$\dot M_{cold}=\dot M_\ast$}
\label{fig2}
\end{figure}

Results of models illustrating the effect of $\eta$ and mass-loading on
the expected cluster temperature are shown in Fig.~\ref{fig2}, and show
that, not surprisingly, both mass-loading and $\eta<1$ lead to lower
cluster temperatures, but also, often, higher X-ray luminosities.

\subsection{X-ray Luminosity Scaling Relations}

Given that the gas temperature within the cluster is comparatively
constant (see Fig.~\ref{fig1}) and hot (for ${\bar V}_\ast\geq 1000\kms$
the temperatures are $>10^7$K, so that the bremsstrahlung region of the
cooling curve is appropriate, i.e.  $\Lambda(T)\propto T^{1/2}$) the
following simple scaling relationships for the X-ray luminosity of
clusters can be derived (ignoring for the time being the effects of
$\eta<1$ and mass-loading).

The gas density within the cluster, $n\propto {\dot M_\ast}/(R_c^2 {\bar
V}_\ast)$, so that if we define a cluster wind scaling parameter
$X_{cl}$ such that

\begin{equation}
X_{cl} = \frac{\dot M_\ast^2}{R_c {{\bar V}_\ast}}
\label{eqn10}
\end{equation}

\noindent then from eqn.~(\ref{eqn8}) $L_X \propto X_{cl}$. We shall
calculate $X_{cl}$ using the natural units, that is $\dot M$ in
$\msunyr$, $R_c$ in pc and ${\bar V}_\ast$ in $\kms$, so that for a
model with $\dot M_\ast=10^{-4}\msunyr$, ${\bar V}_\ast=2000\kms$ and
$R_c=1$pc, then $X_{cl}=5\times 10^{-12}$.

At smaller velocities the deviations from the $T^{1/2}$ scaling for the
cooling curve will alter things. For temperatures in the range of
$10^{6.2}\leq T\leq 10^{6.5}$~K, then roughly $\Lambda(T)\propto
T^{-0.6}$ and the corresponding expression for the variation of the
X-ray luminosity is

\begin{equation}
L_X \propto \frac{\dot M_\ast^2}{R_c {{\bar V}_\ast}^{3.2}} \ .
\end{equation}

The scaling effect of variation in $\eta$ on the cluster properties can
also be derived. A reduction in $\eta$ effectively reduces the amount of
thermal energy available to the cluster as a whole. For the same amount
of mass-loss this will reduce the temperature, with $T_{cl}\propto
\eta$, but also reduce the velocity of the cluster wind and increase the
density.

\begin{figure}
\vspace{7cm}
\includegraphics{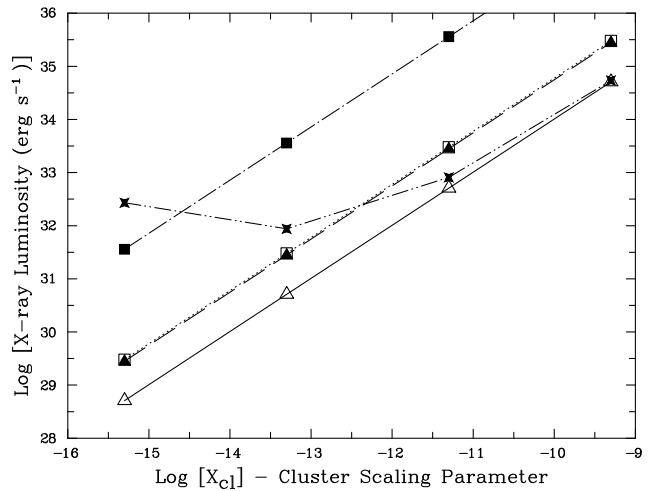} 
\caption{The theoretical cluster X-ray luminosity versus the cluster
wind scaling parameter $X_{cl}={\dot M^2}/(R_c {\bar V}_\ast)$.  Results
are shown for several different sets of models. The cluster radius is
always 1pc and ${\bar V}_\ast=2000\kms$. The open triangles (and solid
line) are for the standard models, which have no mass injection ($\dot
M_{cold}=0$) and $\eta=1$, and stellar mass-loss rates of $\dot
M_\ast=10^{-6}, 10^{-5}, 10^{-4}$ and $10^{-3}\msunyr$. The filled
triangles (dashed line) have the same parameters except $\eta=0.1$. 
The open squares (dashed line) have $\eta=1$ and $\dot M_{cold}=\dot
M_\ast$ (and have very similar results to the previous models). The
filled squares (dot-dashed line) have $\eta=1$ and $\dot M_{cold}=10
\dot M_\ast$.  The crosses (dot-dot-dashed lines) have $\eta=1$ and
$\dot M_{cold}=2\times 10^{-5}\msunyr$.}
\label{fig3}
\end{figure}

For the bremsstrahlung dominant case ($\Lambda(T)\propto T^{1/2}$), the
density $n\propto \dot M/(R_c^2 {{\bar V}_\ast})$ and ${{\bar
V}_\ast}^2\propto \eta$, so that here $L_X\propto \eta^{-0.5}$. In the
case of $\Lambda(T)\propto T^{-0.6}$, then $L_X\propto
\eta^{-1.6}$. Somewhat counter-intuitively then, a reduction in $\eta$
leads to a rise in X-ray luminosity, due to the dominant effect of the
increasing density. Of course, for $\eta\ll 1$ the emission will not be
at X-ray energies and $L_X$ will decrease. Results for the variation of
the cluster X-ray luminosity versus this cluster scaling parameter are
shown in Fig.~\ref{fig3}, where models for a range of cluster wind
scaling parameters $X_{cl}$ are shown.

The effect of mass-loading is similar to the effect of lower values of
$\eta$ -- both lead to a drop in the average available energy per
particle. This is reflected in that models with $\eta=0.1$ have almost
identical results to a model with $\dot M_{cold}=\dot M_\ast$.
Mass-loading leads to a decrease in temperature and an increase in
density. Model results are also shown in Fig.~\ref{fig3}. Of note is
that the fact that we get a very different relationship between $L_X$
and $X_{cl}$ when the mass-loading injection is constant rather than a
constant multiple of $\dot M_\ast$.

\begin{figure}
\vspace{7cm}
\includegraphics{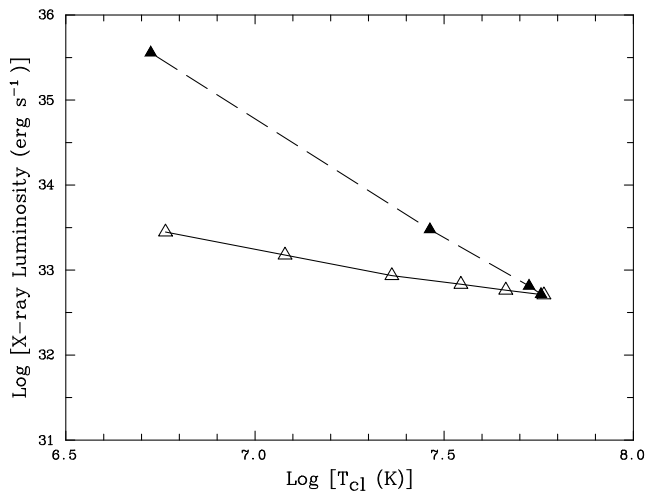} 
\caption{The theoretical stellar cluster X-ray luminosity-temperature
relationship. Shown are model results for the mean cluster temperature
and cluster luminosity for varying $\eta$ and the contribution of cold
gas. The standard cluster model has $\dot M_\ast=10^{-4}\msunyr$,
${\bar V}_\ast=2000\kms$ and $R_c=1$pc.  The models represented by open
triangles and marked with a solid line are for varying $\eta$, with
models with $\eta=1.0,0.8,0.6,0.4,0.2$ and 0.1, with decreasing
temperature corresponding to decreasing $\eta$. The models represented by
filled triangles and marked with a dashed line are for variations with
the amount of cold material injected $\dot M_{cold}$ and have
$\eta=1.0$. The model results are for $\dot M_{cold}=10^{-6}, 10^{-5},
10^{-4}$ and $10^{-3}\msunyr$, with decreasing temperature corresponding
to increasing $\dot M_{cold}$.}
\label{fig4}
\end{figure}

Disentangling the combined action of lower values of $\eta$ and
mass-loading in an individual cluster may be difficult, but in a
statistical sample the relative influence of both effects may be
apparent in star cluster X-ray luminosity--temperature plots. This is
illustrated in Fig.~\ref{fig4}, where for these simple models, different
slopes in the cluster X-ray luminosity-temperature relationship are
found when the effects of $\eta$ and mass-loading are isolated and
varied in a systematic manner. Whether in reality the two effects can be
separated so cleanly is unclear. These models have also assumed a single
cluster core radius. More realistic models, with a range of parameters
will show more scatter.

\begin{table*}
\caption{The observed and predicted properties of the cluster wind from
the 5 star clusters under consideration. The values for ${\dot M}$ and
${\bar V}_\ast$ are obtained using eqns.~\ref{eqn3} and~\ref{eqn5} and
are described more in the text. The predicted temperature, $kT_0$, is
calculated using eqn.~\ref{eqn9}. The expression for the cluster wind
scaling parameter, $X_{cl}$, is given by eqn.~\ref{eqn10}.  The sources
for observed values for $L_X$ and $kT_X$ for each cluster are described
in the text too.}
\begin{center}
\begin{tabular}{lcccccccc} \hline
& \multicolumn{5}{c}{Predicted Values} &\phantom{111} & 
\multicolumn{2}{c}{Observed values}\\ \cline{2-6}\cline{8-9}
&&&&&&&&\\
Cluster & ${\dot M_\ast}$ & ${\bar V}_\ast$ & $R_c$ &  $X_{cl}$ 
& Predicted $kT_0$ && Observed $kT_X$ & $L_X$ \\ 
 &($\msunyr$) & ($\kms$) & (pc) & & (keV) && (keV) & (\ergs) \\ \hline
NGC\,3603 & $2.3\times 10^{-4}$ & 2844 & 2.8 & $6.6\times10^{-12}$ & 
10.1 && 3.7 & $2.2\times 10^{34}$ \\ 
R136      & $2.6\times 10^{-4}$ & 2110 & 2.0 & $1.6\times10^{-11}$ & 
5.6  && 2.1 & $5.5\times 10^{34}$ \\
NGC\,346  & $1.7\times 10^{-5}$ & 2282 & 2.0 & $6.4\times 10^{-14}$ & 
6.5  && 1.0 &  $1.5\times 10^{34}$ \\
Rosette   & $2.5\times 10^{-6}$ & 2173 & 4.6 & $6.3\times 10^{-16}$ & 
5.9  && 0.6 & $8.0\times 10^{32}$ \\
Arches    & $7.3\times 10^{-4}$ & 2810 & 0.2 & $9.5\times 10^{-10}$ & 
9.9  && 0.8,6.4 & $5.0\times 10^{35}$ \\ \hline
\end{tabular}
\end{center}
\label{tab2}
\end{table*}

\section{The Sample of Nearby Star Clusters}

SSCs and other compact star clusters are relatively small objects with a
typical size of a few pc, so even with the arcsecond resolution of {\sl
Chandra} only those that are nearby can be studied in any detail. In
this study {\sl Chandra} data for R136, NGC\,3603, NGC\,346, the Rosette
Nebula and the Arches cluster are used. The definition of an SSC is
rather vague and certainly the Rosette Nebula does not qualify. However,
although it is a much smaller cluster the physics should be the same and
so it is included. In this section we describe both the relevant
properties of the stellar components of the cluster and the diffuse
X-ray properties of the X-ray emission as seen with the {\sl Chandra}
satellite. In some cases results from the literature are used, while in
others an analysis of the data has been performed.

\subsection{NGC\,3603}

NGC\,3603 is the most massive visible H\small{II} region in our galaxy,
and is commonly regarded as a Galactic analogue for extragalactic
SSCs. It contains a very dense concentration of stars of ages of $\sim
2-3$Myr, including 3 WR stars in its core. In many ways it is similar to
R136 in 30~Dor but without the surrounding cluster halo (Moffat \etal\
1994). We assume a distance of 7~kpc.

The characteristics of the massive stars in NGC\,3603 have been studied
by Crowther \& Dessart (1998), and using this data and Kudritzki \& Puls
(2000), the mass-loss rates and terminal velocity data for 44 stars
within 2.8pc of the centre of the cluster have been tabulated. These
stars will dominate the stellar mass and energy loss into the cluster.
Based on these results, the total stellar mass-loss rate for NGC\,3603
is estimated as $\dot M_\ast=2.3\times 10^{-4}\msunyr$ and the mean
weighted terminal velocity as ${\bar V}_\ast=2844\kms$. The cluster core
radius is taken to be 2.8pc (Moffat \etal\ 1994).

Results from a 50ksec ACIS-I {\sl Chandra} observation of NGC\,3603 have
already been presented in Moffat \etal\ (2002), and these data show
strong evidence of a diffuse thermal component, probably associated with
a cluster wind. To determine the spectral characteristics of the diffuse
emission, point sources are excluded and the diffuse spectrum
extracted. The spectrum has been fitted with an absorbed {\it mekal}
model ({\it wabs*mekal}), with $N_H=0.67\times 10^{22}\cmsq$,
$kT_X=3.7$keV, and the absorption corrected luminosity is $L_X=2.2\times
10^{34}\ergs$ (see Table~\ref{tab2}).

\subsection{R136 in 30 Doradus}

R136 is the central object of 30 Doradus in the LMC, and is regarded as
the closest example of an SSC, and we assume a distance of 50~kpc
(cf. Feigelson 2001). The spectral types and mass-loss rates of 39 stars
in R136 were obtained from a study by Crowther \& Dessart (1998), and the
corresponding wind velocities from Kudritzki \& Puls (2000). The cluster
mass-loss rate is estimated to be $\dot M_\ast=2.6\times 10^{-4}\msunyr$
and the mean weighted terminal velocity as ${\bar V}_\ast=2110\kms$. The
cluster core radius is assumed to be 2pc (Crowther \& Dessart 1998).

The X-ray data of R136 was obtained from the {\sl Chandra} archive and
consists of a 28ksec ACIS-I observation of 30 Doradus. The point sources
within a region of radius $16{''}$ centred on the cluster were removed
as for NGC\,3603, and a spectrum was extracted from the remaining
emission, and fitted again using an absorbed {\it mekal} model.  The
best fit spectrum to this emission was an absorbed {\it mekal} model,
with $N_H=0.42\times 10^{22}\cmsq$, $kT_X=2.1$keV, and absorption
corrected $L_X=5.5\times 10^{34}\ergs$ (see Table~\ref{tab2}).

\subsection{NGC\,346}

NGC\,346 is the largest star formation region of the SMC, and contains
the majority of the O stars in the SMC (Massey, Parker \& Garmany
1989). We assume a a distance of 59\,kpc (Mathewson, Ford \& Visvanathan
1986).  The spectral types and magnitudes of the stars in NGC\,346 were
obtained from Massey \etal\ (1989). Using a reddening correction of
$E(B-V)=0.14$, mass-loss rates were calculated from the expression in
Howarth \& Prinja (1989), accounting for the effects of lower
metallicity on stellar mass-loss rates, with $\dot M(Z)\propto
(Z/Z_\odot)^{0.5}$ (cf Kudritzki \& Puls 2000 and using a value for the
SMC metallicity of $Z_\odot/5$).  The cluster mass-loss
rate for NGC\,346 is estimated to be $\dot M_\ast=1.7\times
10^{-5}\msunyr$ and the mean weighted terminal velocity as ${\bar
V}_\ast=2282\kms$. The cluster core radius is taken to be 2pc (Massey
\etal\ 1989).

{\sl Chandra} observed NGC\,346 for a total of 100ksec in May 2001. The
cluster lies close to a chip gaps of the ACIS-I instrument, but diffuse
emission from the cluster is still detected. A more detailed analysis of
the {\sl Chandra} data of the NGC\,346 region can be found in Naz\'e
\etal\ (2002a; 2002b). The extracted spectrum was best fit with an
absorbed {\it mekal} model, with $N_H=0.42\times 10^{22}\cmsq$,
$kT_X=1.0$keV, and $L_X=1.5\times 10^{34}\ergs$ (absorption corrected,
see Table~\ref{tab2}).\\

\begin{figure*}
\vspace*{7cm}
\includegraphics{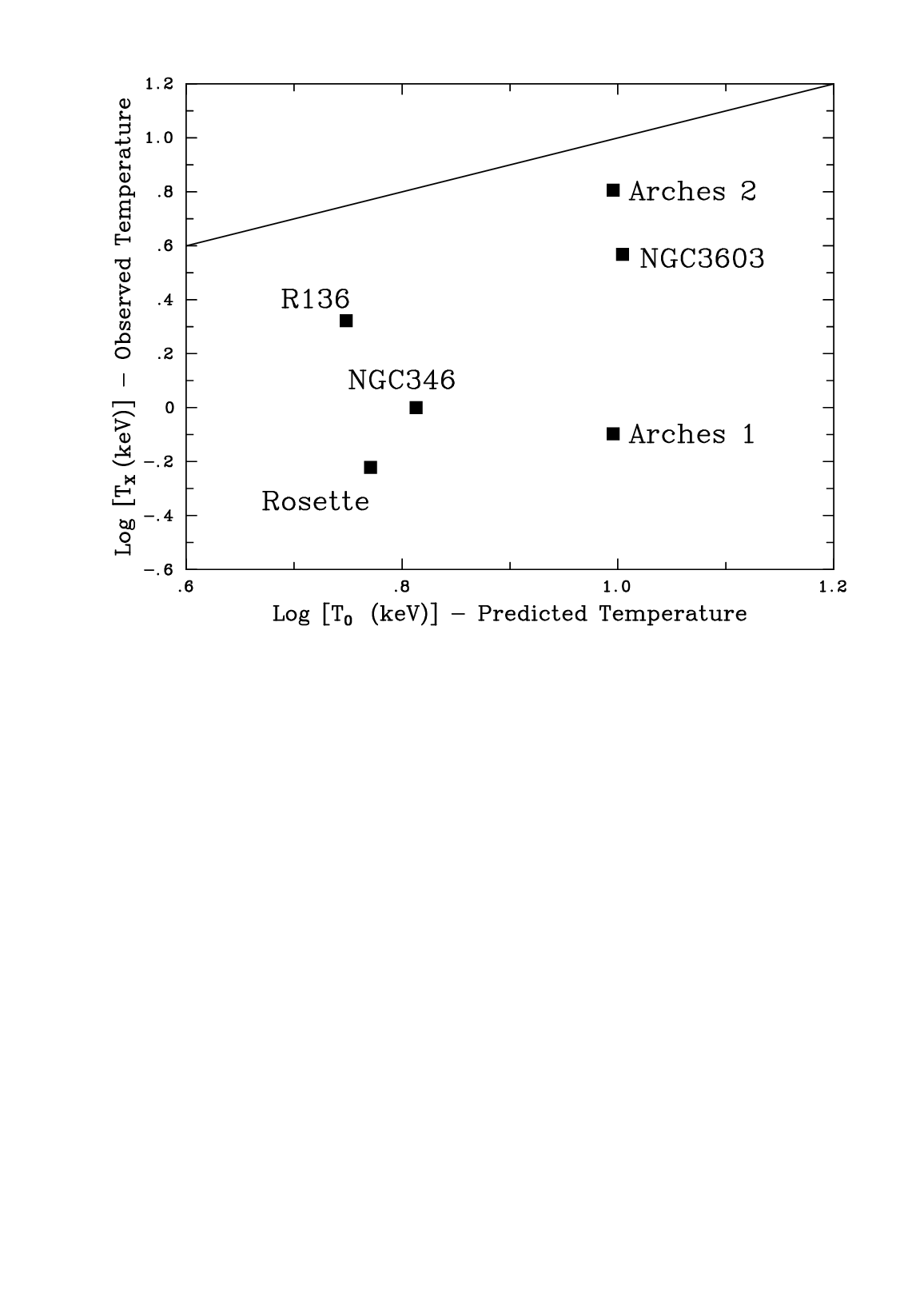} 
\includegraphics{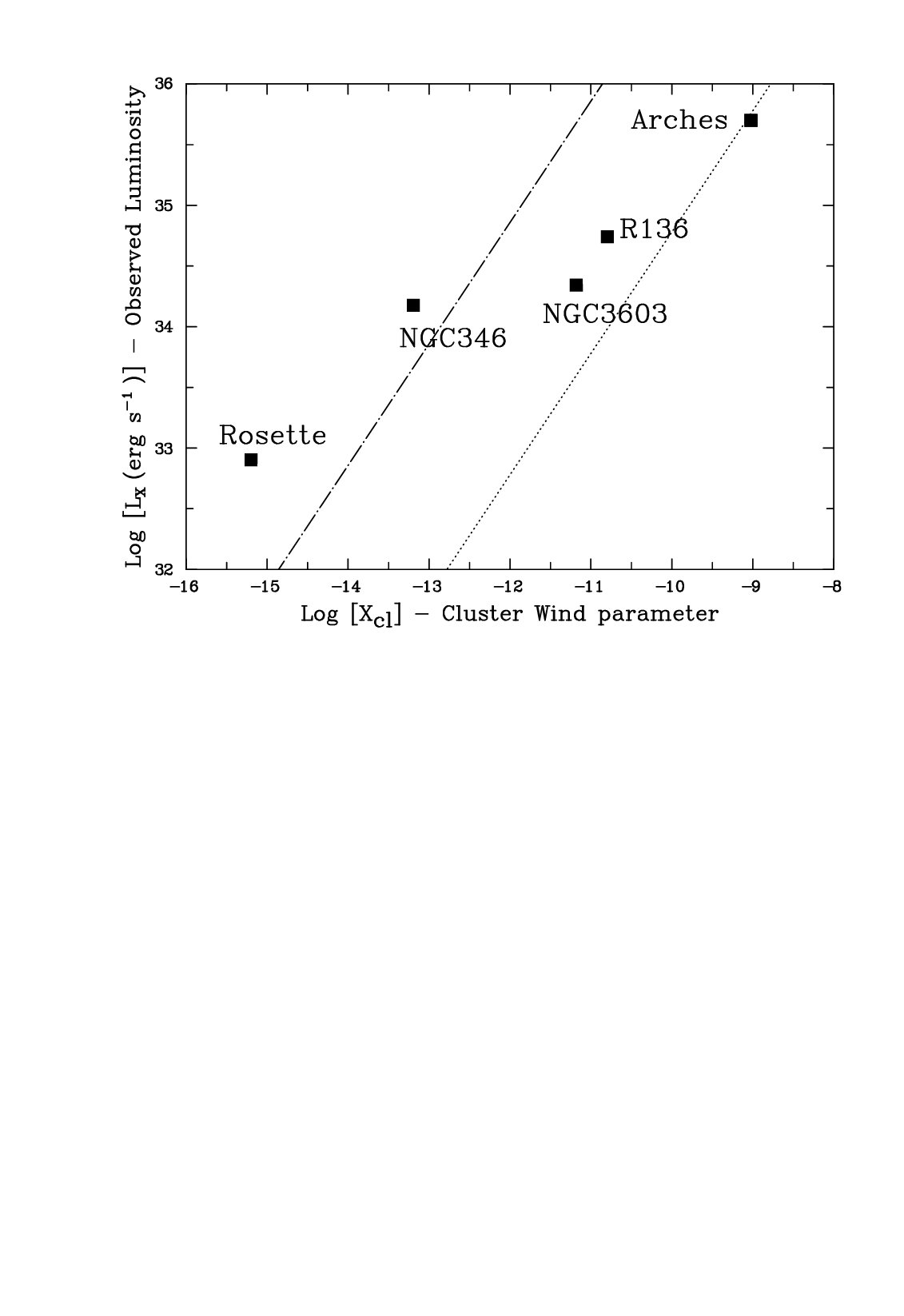} 
\caption{Left: The fitted X-ray temperature of the diffuse X-ray
emission of the sample of star clusters versus the  predicted X-ray
temperature calculated from the model described in \S~2. The solid line
depicts equality, that is extremely efficient conversion of
the kinetic energy in the winds of the component stars to thermal
energy. Two values of observed temperature for the Arches cluster are
plotted (see text for details). Right: The observed X-ray luminosity
$L_X$ versus the theoretical cluster wind parameter $X_{cl}$
(eqn.\ref{eqn10}) for the
sample clusters. Also plotted are two results from two models shown in
Fig.~\ref{fig3}, namely models with $\eta=1$ and $\dot M_{cold}=\dot
M_\ast$ (dotted line) and $\dot M_{cold}=10 \dot M_\ast$ (dot-dashed
line). } 
\label{fig5}
\end{figure*}

\subsection{The Rosette Nebula, NGC\,2237}

The Rosette Nebula (NGC\,2237) contains an open cluster, NGC\,2244, and
is one of the more massive diffuse nebulae. It is situated at a distance
of 1.5\,kpc (Gregorio-Hetem \etal\ 1998). The cluster wind parameters of
the Rosette Nebula were calculated using the properties of the 26 O, B
and A stars lying within 4.6pc of the cluster centre, which corresponds
to the rough extent of the X-ray emission.

Spectral types were obtained from {\sl SIMBAD}, and the corresponding
wind velocities from Kudritzki \& Puls (2000). Mass-loss rates were
evaluated using the relationship of Howarth \& Prinja (1989), assuming
$E(B-V)=0.49$ (Massey, Johnson, DeGioia-Eastwood 1995). The resulting
values for the Rosette Nebula are $\dot M_\ast=2.5\times 10^{-6}\msunyr$
and ${\bar V}_\ast=2173\kms$. The cluster core radius is assumed to be
4.6pc (see above).

{\sl Chandra} observations of this region found a large number of point
sources. On subtraction of these point sources, a weak diffuse component
was discovered (see Montmerle \etal\ 2002; Townsley \etal\ 2002, in
preparation). Fitting this with an absorbed Raymond-Smith model, the
best fit parameters for the diffuse emission were $N_H=7\times
10^{21}\cmsq$, $kT_X=0.6$keV and the absorption corrected luminosity was
$L_X=8\times 10^{32}\ergs$ in the $0.5-2$keV waveband.\\

\subsection{The Arches Cluster}

The Arches cluster is located at a projected distance of $\sim 50$pc
from the Galactic centre (Cotera \etal\ 1996; Portegies-Zwart 2001), and
contains $\sim 120$ stars which have masses greater than $20\msun$
(Serabyn, Shupe \& Figer 1998). It is one of the densest star clusters
known in the local group. Blum \etal\ (2001) have presented a $2\mu$m
study of the stellar population of the Arches cluster, and inferred an
age for the cluster of between 2 and 4.5Myr.  We assume that the cluster
core radius is  0.2pc and that the cluster is at a distance of 8kpc
(Figer \etal\ 1999b). 

Due to the lack of comprehensive data on the stars in the Arches
cluster, we have adopted the simpler approach of calculating $\dot
M_\ast$ and ${\bar V}_\ast$ using the {\sl Starburst99}
models\footnote{Available at
http://www.stsci.edu/science/starburst99}. Based on the results of Figer
\etal\ (1999a), for the Arches cluster we assume a top-heavy IMF (with
$\alpha=1.6)$, solar abundances, high mass-loss stellar evolution tracks
and a total stellar mass of $1.1\times 10^{4}M_\odot$ (for stars in the
mass range $1-100M_\odot$), and the appropriate values are taken for
values in the range $t\sim 2-4.5$Myr. For times $>$3.5Myr mass and
energy injection from SN begin to play a role.  For the Arches cluster,
in this time span the derived values of the mean weighted terminal
velocity are in the range ${\bar V}_\ast=1800-2810\kms$, and the total
stellar mass-loss is in the range $\dot M_\ast=(2.0-9.5)\times
10^{-4}\msunyr$. The corresponding large range in the cluster wind
parameter is $X_{cl}= (0.85-25.0)\times 10^{-10}$.
 
From the {\sl Starburst99} simulations it is interesting to note that
the value of ${\bar V}_\ast \geq 1500\kms$ for all times of relevance to
young clusters. In the {\sl Starburst99} models only single stars are
included and the energy injection rate from SN tails off at $t\sim
35$Myr (in this model). In models including binaries, mass and energy
injection from SN will continue to longer times due to the processes of
mass transfer and binary evolution.

For convenience, we will adopt the values at a time of 4Myr for the
Arches cluster, with ${\bar V}_\ast=2810\kms$, and $\dot
M_\ast=7.3\times 10^{-4}\msunyr$.  These values are of course rather
uncertain, and but for instance we note that direct radio detections of
a number of the stars in the Arches cluster by Lang, Goss \& Rodr\'iguez
(2001) might suggest a broadly similar value for the total mass-loss
rate from the cluster. However, in their simulation of the X-ray
properties of the Arches cluster, Raga \etal\ (2001) assumed a somewhat
larger integrated mass-loss rate (60 stars each with an individual
mass-loss rate of $10^{-4}\msunyr$) but a lower mean terminal velocity
(the wind of each star was assumed to have a terminal velocity of
$1500\kms$). The calculations of Raga \etal\ (2001) did not include
additional mass-loading and in the following section we shall see that
to reproduce the X-ray properties of the Arches (and other) clusters we
find that we do need to include mass-loading (or a lower $\eta$), which
has a similar effect as an enhanced mass-loss rate and reduced terminal
velocity.

Results from a 51ksec ACIS-I {\sl Chandra} observation of the Arches
cluster have been presented by Yusef-Zadeh \etal\ (2002) and show three
main emission regions probably associated with the Arches cluster
(designated A1--A3 by Yusef-Zadeh \etal\ 2002). Two of these components
are compact (A1 and A2) and one coincides with the core of the cluster
(A1). The third component (A3) is more extended and part of it may be
due to the cluster wind and part may be associated with fluorescent
emission from a molecular cloud (C. Lang, private
communication). Yusef-Zadeh \etal\ (2002) fitted the spectrum of this
component with an additional line feature at 6.4keV.  The total X-ray
luminosity ($0.2-10$keV) from the 3 components is $5\times
10^{35}\ergs$, with component A1 being the most luminous.  The spectra
are fitted with a 2 component model, with ${kT_1} \sim 0.8$keV and
${kT_2}\sim 6.4$keV for the A1 component, and we shall use both these
temperature values in the analysis, but only the overall luminosity. The
two compact X-ray sources could be genuine point sources, associated
with colliding stellar winds, or at least one of them could also be the
unresolved core of a cluster wind.\\

\section{Results: Comparison of Theory and Observation}

From the theory developed in \S~2 and the observational results that we
have for the small number of clusters, presented in \S~3 we can now make
some comparisons. In Table~\ref{tab2} we summarise the theoretical and
observational data for the sample clusters.

Fig.~\ref{fig5} (left panel) shows a plot of the observed X-ray
temperatures from the {\sl Chandra} observations versus the
theoretically predicted wind temperatures for each cluster
(specifically, values of $T_0$ are plotted, calculated using
eqn.~\ref{eqn9}). It is clear that in all cases bar one the observed
temperature is significantly lower than that the value predicted from
the stellar population in the cluster. The exception is the case of the
hotter component of the Arches cluster, where the discrepancy is much
less. In the case of the Arches though, it should be remembered that the
best-fit model to the X-ray spectrum had two components and the 2nd
component is also plotted and is much more in line with the other
clusters. In fact, the cooler component (${kT_1}\sim 0.8$keV) is more
luminous than the hotter component and it may have been better to use
that value. Also, R136 is rather more distant than the rest and the
possibility of unresolved point sources contaminating the spectrum is
more. If we were to ignore both the Arches cluster and R136 (of course,
leaving only 3 data points) then a more coherent picture may be emerging
of consistently lower observed temperatures.  As has been noted before,
such a reduction in observed temperature is a natural consequence of
$\eta<1$ or that significant mass-loading is occurring.

In \S~2 the cluster wind model was also used to make predictions
concerning the scaling of the cluster X-ray luminosity versus the
cluster wind scaling parameter $X_{cl}$ (see eqn.~\ref{eqn10}). The
observed correlation between the measured $L_X$ and $X_{cl}$ is shown in
the right panel of Fig.~\ref{fig5}.  The most notable feature is that
compared to the standard model in Fig.~\ref{fig3} the clusters are all
overluminous by over an order of magnitude. The second feature is that
there does seem to be a clear correlation between $L_X$ and $X_{cl}$,
not necessarily linear, but apparently monotonic. For comparison,
results from model calculations, already shown in Fig.~\ref{fig3},
illustrate that the effects of mass-loading can roughly reproduce the
X-ray luminosities of the cluster, with $\dot M_{cold}=1-10 \dot
M_\ast$. Similarly, very low values of $\eta$ could yield the same
result (with $\eta\sim 0.01$), though this would be in conflict with the
observed temperatures.

It could be argued that this correlation is partly a consequence of the
simple fact that we would expect more massive clusters to be more X-ray
luminous, and undoubtedly the value of $X_{cl}$ is most dependent on the
total stellar mass-loss rate of the cluster ($\dot M_\ast$) and this
will in turn scale with stellar mass. Larger clusters may also have
larger populations of unresolved point sources making them appear more
X-ray luminous. With the very limited sample we have it is difficult to
go further than merely noting that there is a correlation between $L_X$
and $X_{cl}$.  The Rosette nebula appears to lie at a somewhat higher
luminosity than its value of $X_{cl}$ would imply. This could imply a
larger comparative fraction of mass-loading as compared to the other
clusters (see Fig.~\ref{fig3}). The low observed temperature, as
compared to the expected value, would tally with this.

\begin{figure}
\vspace*{7cm}
\includegraphics{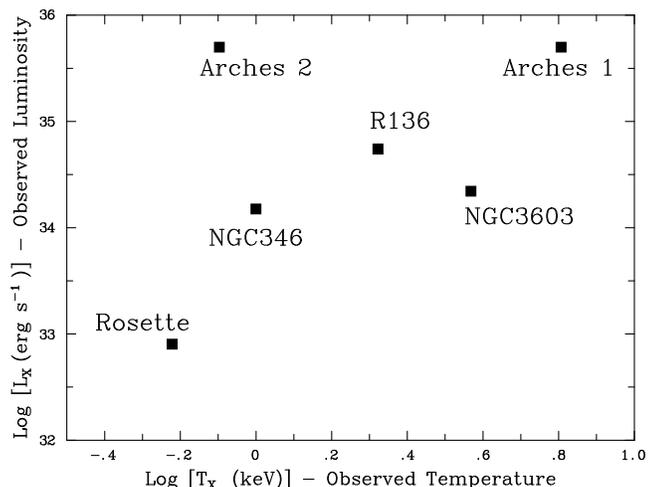} 
\caption{A stellar cluster X-ray luminosity-temperature diagram for the
observed clusters, see text for more details. Note that the two values
plotted for the Arches cluster correspond to the two fitted components
(see text for details).} 
\label{fig6}
\end{figure}

In Fig.~\ref{fig3} we postulated that it may be possible to discriminate
between the effects of a low thermalization efficiency and mass-loading
by means of a stellar cluster X-ray luminosity temperature diagram. In
Fig.~\ref{fig6} we show a plot of the observed values for the cluster
sample. The model predicted that the luminosity and cluster temperature
should be inversely correlated (though with different slopes depending
on whether mass-loading or low thermalization efficiency was
dominant). From Fig.~\ref{fig6} we can see clearly that the data suggest
that X-ray temperature and luminosity are probably not well correlated.\\

\section{Summary and Conclusions}

In this paper we have presented a comparison between the theory of
outflows from young star clusters in a cluster wind and X-ray
observations of diffuse X-ray emission from these clusters. The high
spatial resolution of {\sl Chandra} is necessary to begin to disentangle
the diffuse emission from point source emission. However, even with {\sl
Chandra} problems remain as to being sure that the diffuse emission is
genuinely diffuse and associated with the cluster wind. The major
problem remains the extremely small sample size. More observations will
alleviate this problem to some extent, but the objects discussed here
represent the best examples of such a cluster wind.

This analysis has thrown up some interesting results, the diffuse X-ray
luminosity of the star clusters in the sample is correlated with the
cluster wind scaling parameter $X_{cl}= {\dot M_\ast^2}/({R_c {{\bar
V}_\ast}})$ as predicted, but that the observed X-ray temperature is not
well correlated with the predicted X-ray temperature (though a better
correlation may be masked by the nature of the data). It is also clear
that from these data it is very unclear as to what is going on in the
clusters; is there a low thermalization efficiency or is mass-loading the
dominant mechanism?

In this paper we have put together the results from several {\sl
Chandra} observations of nearby massive young star clusters to
investigate the cluster wind. It is clear that many more observations
will be required to take the study further, to make a clearer connection
between the theory and observations of cluster winds. Such observations
will be necessary to really understand the impact that star clusters
will have on the interstellar medium of galaxies of all shapes and
sizes.

\end{document}